%% file: apstemplate.tex
\documentclass[aps,amsmath,amssymb,reprint]{revtex4-2}
\usepackage{url}
\usepackage{graphicx}
\usepackage[colorlinks,linkcolor=blue,citecolor=blue]{hyperref}

\newcommand{\xim}{\Xi^{-}}
\newcommand{\xip}{\bar{\Xi}^{+}}
\newcommand{\lambar}{\bar{\Lambda}}
\newcommand{\XXb}{\Xi^{-}\bar{\Xi}^{+}}
\newcommand{\LLb}{\Lambda\bar{\Lambda}}

\newcommand{\jpsi}{J/\psi}
\newcommand{\chifc}{\chi^2_{\rm 4C}}
\newcommand{\EE}{e^+e^-}
\newcommand{\BESIIIorcid}[1]{\href{https://orcid.org/#1}{\hspace*{0.1em}\raisebox{-0.45ex}{\includegraphics[width=1em]{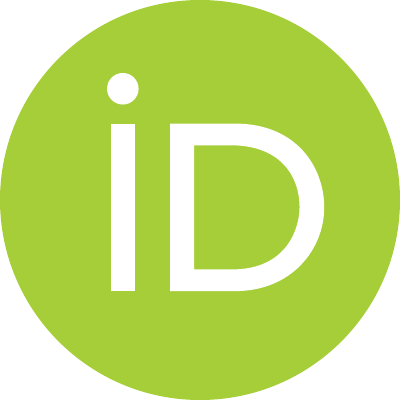}}}} 

\begin{document}

\title{Precise Measurement of Matter-Antimatter Asymmetry with Entangled Hyperon Antihyperon Pairs}
\author{BESIII Collaboration}
\thanks{Full author list given at the end of the Letter.}

\begin{abstract}
A search for $CP$ violation with an entangled system of $\Xi^-\bar\Xi^+$ pairs is performed, using $(10,087\pm44)\times10^{6}$ $J/\psi$ events collected with the BESIII experiment. A nine-dimensional helicity amplitude is used to fit $\EE\to\jpsi\to\XXb$ and its subsequent decays. The $\xim$ and $\xip$ decay parameters are determined with higher precision compared to the best results reported so far. Furthermore, the strong phase difference, $(\delta_P-\delta_S)=(0.3\pm1.2\pm0.2)\times10^{-2}~\text{rad}$, and the weak phase difference, $(\xi_P-\xi_S)=(-0.2\pm1.2\pm0.1)\times10^{-2}~\text{rad}$, are directly determined. These are the most precise measurements to date. The $CP$ asymmetry observables $A_{CP}^{\Xi}=(-7.8\pm4.8\pm0.8)\times10^{-3}$ and $\Delta\phi_{CP}^{\Xi}=(0.6\pm5.1\pm0.2)\times10^{-3}~\text{rad}$ are determined, which are consistent with $CP$ conservation.
In addition, independent measurements of the $\Lambda$ decay parameter and $CP$ asymmetry $A^{\Lambda}_{CP}=(-2.9\pm4.3\pm0.7)\times10^{-3}$ are also obtained, which are in agreement with the previous measurements, but with much improved precision.

\end{abstract}

\maketitle

One of the main unresolved puzzles in fundamental physics is to explain how the vast abundance of matter over antimatter in the universe was generated. The most viable explanation is a dynamic mechanism, Sakharov’s theory of \textit{baryogenesis}~\cite{Sakharov:1967dj}. For such a mechanism to leave a lasting imprint, it must meet a few essential criteria, one of which requires physics processes that violate the combined charge conjugation and parity ($CP$) symmetry. So far, $CP$ violation has only been discovered in weak flavor-changing quark transitions.  Within the standard model (SM), it is accounted for through a single irreducible complex phase in the Cabibbo-Kobayashi-Maskawa matrix \cite{Cabibbo:1963yz, Kobayashi:1973fv}. However, this mechanism is insufficient to explain the observed matter-antimatter asymmetry of our universe~\cite{Bernreuther:2002uj, Canetti:2012zc}. Experimentally, $CP$ violation has been well established in strange, beauty, and charm meson decays~\cite{Christenson:1964fg, Aubert:2001nu, Abe:2001xe, Aaij:2019kcg}. Only recently, the LHCb collaboration observed $CP$ violation in a beauty baryon decay, $\Lambda_b^0\to pK^-\pi^+\pi^-$~\cite{LHCb:2025ray}. 
Since baryons have an additional spin degree of freedom, precise $CP$ measurements test the SM in a complementary way. For nonleptonic two-body weak decays of hyperons, direct $CP$ violation can be compared to that of kaon nonleptonic weak decays~\cite{Tandean:2003fr, Salone:2022lpt}. To observe $CP$ violation, there should be two or more weak amplitudes which can interfere with each other. 
For kaons, the interference effects, that give rise to $CP$ violation, come from the two-pion isospin states $I_{\pi\pi}=0, 2$. For the $\Delta S=1$ transition in $\xim \to \Lambda \pi^-$, 
the two interfering amplitudes come from the parity-odd ($S$-wave) and parity-even ($P$-wave) decay amplitudes.
For the $L=S, P$ amplitudes, the strong and weak phase components are denoted as $\delta_L$ and $\xi_L$, respectively. The full amplitude is given by $L= \sum_{j} L_j \exp (i\delta_{j}^L + i\xi_{j}^L)$, where $j$ runs over all possible combinations of $[2I,2\Delta I]$. The coefficient $L_j$ is real, and $I$ and $\Delta I$ represent the isospin of the final state and the change of isospin from the initial to final state, respectively~\cite{Salone:2022lpt, Donoghue:1986hh}. While the strong phases $\delta_L$ do not change sign for the charge-conjugated states, the weak phases $\xi_L$ do. 

\begin{figure}[!htbp] 
	\centering
	\includegraphics[width=0.48\textwidth]{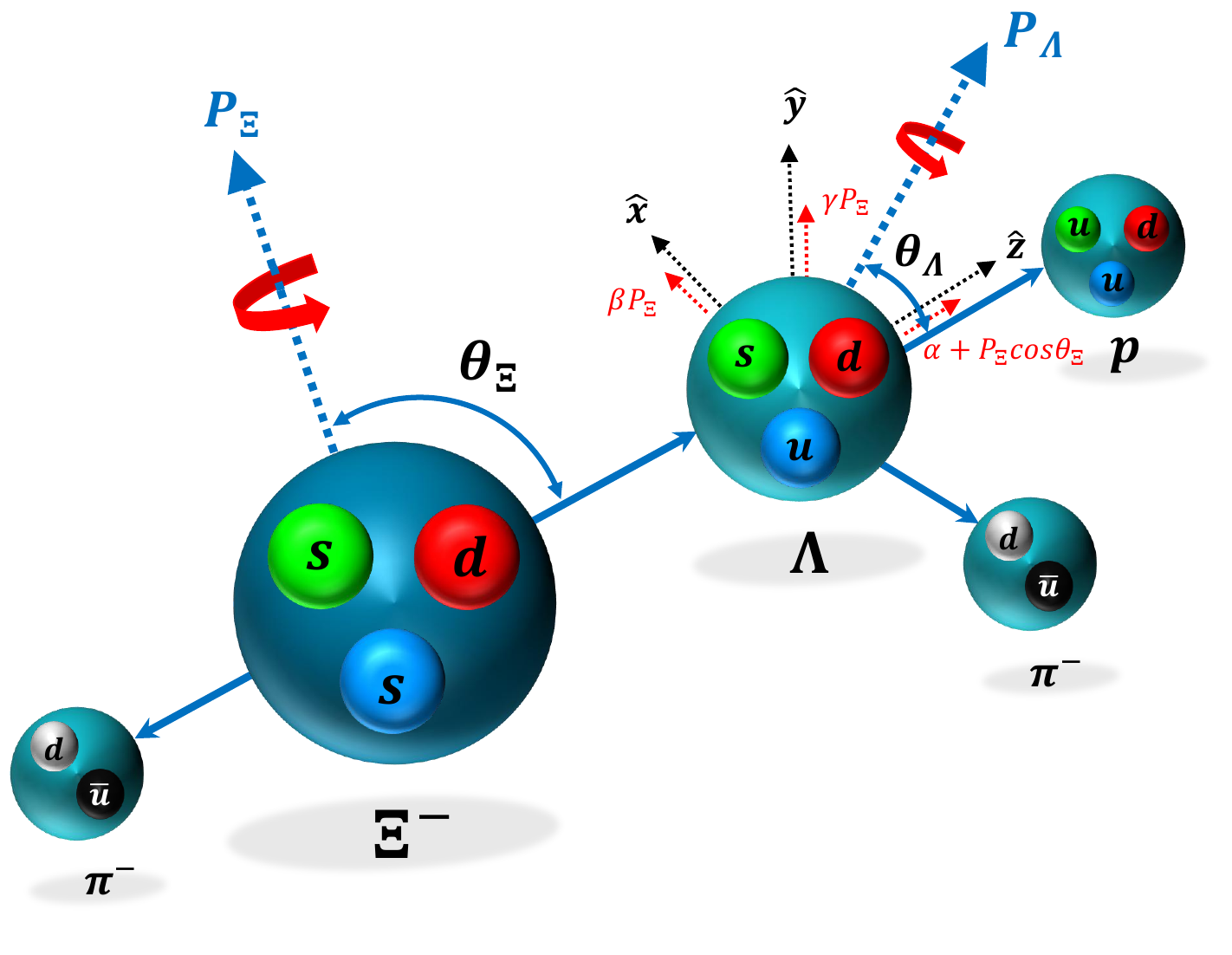} 
        \caption{Illustration of the polarization vectors of $\Lambda$ and $\Xi^-$ hyperons in relation to the decay parameters $\alpha$, $\beta$, and $\gamma$ for the $\Xi^-\to\Lambda\pi^-$ with $\Lambda\to p\pi^-$ process, in the $\Xi^-$ rest frame.}
	\label{fig:orient} 
\end{figure}

For a nonleptonic two-body weak decay, the angular distributions of the final-state particles allow for the extraction of three decay parameters, $\alpha$, $\beta$, and $\gamma$, which are described by the $S$- and $P$-wave amplitudes~\cite{Lee:1957qs}. In addition the parameters $\beta$ and $\gamma$ can also be described via $\alpha$ and the independent phase angle $\phi$, $\beta=\sqrt{1-\alpha^2}\sin{\phi}$ and $\gamma=\sqrt{1-\alpha^2}\cos{\phi}$. The decay parameters are related to the polarization of the decaying system, as shown in Fig.~\ref{fig:orient}. 
In the helicity reference system for the decay $\Xi^-\to \Lambda \pi^-$, the $\hat{\boldsymbol{z}}$ axis is defined by the momentum direction of the final state baryon ($\Lambda$) momentum in the $\Xi^-$ rest frame. The $\hat{\boldsymbol{y}}$ axis is defined as the cross product of the $\hat{\boldsymbol{z}}$ axis and the initial polarization $\boldsymbol{P}_{\Xi}\times\hat{\boldsymbol{z}}$. 
When decaying, the initial $\Xi^-$ polarization, $\boldsymbol{P}_{\Xi}$, is transferred to the final state baryon, both longitudinally and transversely. 
The longitudinal component depends on both the initial polarization and the value of $\alpha_{\Xi}$, while $\phi_{\Xi}$ is directly proportional to $\boldsymbol{P}_{\Xi}$. 
The parameters $\alpha_{\Xi}$ and $\phi_{\Xi}$ are $CP$-odd and can be used to test the $CP$ symmetry. The $\alpha_{\Xi}$, along with the decay parameter of the charge-conjugated process $\bar{\Xi}^{+}\to \bar{\Lambda}\pi^+$ denoted as $\overline{\alpha}_{\Xi}$, can be used to define the $CP$ observable $A_{CP}^{\Xi}$. 
The dominant transition in such processes is $\Delta I=1/2$~\cite{BESIII:2023jhj,Salone:2022lpt}.
At leading order,
\begin{equation}
     A_{CP} = \frac{\alpha + \overline\alpha}{\alpha - \overline\alpha} = - \tan(\delta_P - \delta_S)\tan(\xi_P - \xi_S). 
     \label{eq:cp}
\end{equation}

\noindent
The strong phase difference $(\delta_P-\delta_S)$ signifies the final-state interaction between the daughter baryon and pion,
while the weak phase difference, $(\xi_P - \xi_S)$, denotes the strength of the $CP$ violation contribution. Since the strong phases are small~\cite{BESIII:2021ypr,Salone:2022lpt}, any $CP$ violation in $A_{CP}^{\Xi}$ would be suppressed. However, with $\phi_\Xi$ and its charge conjugate, $\overline{\phi}_\Xi$, it becomes possible to disentangle and determine both the strong and weak phase differences via
\begin{equation}
(\delta_P - \delta_S)_{\Delta I} = \frac{\sqrt{1-\langle\alpha_\Xi\rangle^2}}{\langle\alpha_\Xi\rangle} \langle\phi_\Xi\rangle,
\label{eq:strP}
\end{equation}
\begin{equation}
(\xi_P - \xi_S)_{\Delta I} = \frac{\sqrt{1-\langle\alpha_\Xi\rangle^2}}{\langle\alpha_\Xi\rangle} \Delta\phi_{CP}^{\Xi}. 
\label{eq:weakP}
\end{equation}
Here $\langle\alpha_\Xi\rangle$ and $\langle\phi_\Xi\rangle$ are the average values, $\langle\alpha_\Xi\rangle=(\alpha_{\Xi}-\overline{\alpha}_{\Xi})/2$ and $\langle\phi_\Xi\rangle=(\phi_{\Xi}-\overline{\phi}_{\Xi})/2$, respectively. The variable $\Delta\phi_{CP}^{\Xi}=(\phi_{\Xi}+\overline{\phi}_{\Xi})/2$ is the second $CP$ observable, which can be determined experimentally. 
In particular, the weak phase differences in hyperon $\Delta S=1$ decays are related to the $\epsilon'/\epsilon$ parameter determined from kaon decays. A general theoretical framework can be constructed to relate potential contributions beyond the SM (BSM), represented by chromomagnetic-penguin operators, across the observables of both hyperon and kaon decays. This formal connection enables a consistent analysis of new physics effects in these distinct but related decay processes. Currently, the constraints are still set by the kaon data, which give the upper limit of the BSM contribution as $(\xi_P-\xi_S)_{\rm BSM} \leq 3.7\times10^{-3}~\text{rad}$ for the $\Xi\to \Lambda \pi$ decay~\cite{Tandean:2003fr, Salone:2022lpt}.

To determine the weak phase difference from a single weak decay chain $\Xi^- \to \Lambda(\to p \pi^-) \pi^-$, the $\Xi^-$ and $\bar{\Xi}^+$ must be significantly polarized. Alternatively, if the $\Xi^-\bar{\Xi}^+$ pairs are produced in a spin-entangled system, as they are for $J/\psi\to\Xi^-\bar{\Xi}^+$ decays, the strong and weak phases are accessible from the $\Xi^- -\bar{\Xi}^+$ spin correlations~\cite{Adlarson:2019jtw, BESIII:2021ypr}. This method was first implemented for the study of $\jpsi\to\XXb$ decays by the BESIII Collaboration, based on $1.3\times 10^9$ $J/\psi$ events, where the weak phase difference was quantified as $(\xi_{P}-\xi_{S})=(1.2\pm3.4\pm0.8)\times10^{-2}~\text{rad}$~\cite{BESIII:2021ypr}. Subsequent determinations of the weak phase difference with BESIII for the  $\jpsi\to\Xi^0[\to\Lambda(\to p\pi^-)\pi^0]\bar{\Xi}^0[\to\bar{\Lambda}(\to \bar{p}\pi^+)\pi^0]$~\cite{Besiii:2023drj} and $\jpsi\to\Xi^-[\to\Lambda(\to p\pi^-)\pi^-]\bar{\Xi}^+[\to\bar{\Lambda}(\to \bar{n}\pi^0)\pi^+]$~\cite{BESIII:2023jhj} processes were performed with 
the same formalism. For all these measurements, the strong phase differences were found to be compatible with zero. 
This stands in contrast to the latest theoretical estimate, $(15.4\pm0.3)\times10^{-2}~\text{rad}$~\cite{Huang:2017bmx}, and to the single non-BESIII experimental result $(10.2\pm3.9)\times 10^{-2}~\text{rad}$, obtained by the Fermilab-based HyperCP experiment~\cite{HyperCP:2004zvh} combined with $\alpha_{\Xi}=-0.376$~\cite{BESIII:2021ypr,Note1}.
The available theoretical predictions of the strong phase difference use different models, such as the N/D method~\cite{Nath:1965iud}, the baryon chiral perturbative theory~\cite{Lu:1994ex, Kamal:1998se, Datta:1998pv, Datta_1995, Tandean:2000dx, Kaiser:2001hr, Barros:2004pw, Huang:2017bmx}, 
or the relativistic chiral unitary approach~\cite{Meissner:2000re, Oller:2005ig, Oller:2006jw, Guo:2012vv}.
However, current theoretical frameworks are still unable to reliably predict the strong phase difference of $\Xi^-\to\Lambda\pi^-$. Here, the experimental results serve as a guide for the theoretical community.\\

In this Letter, we present the most precise determination of the strong and weak phase differences for the $\Xi^-$ hyperon decay. This is done for the $\xim\to\Lambda\pi^-$ decay based on $(10\,087\pm44)\times10^{6}$ $J/\psi$ events collected by the BESIII experiment~\cite{BESIII:2021cxx}. This dataset is approximately eight times larger than the original measurement~\cite{BESIII:2021ypr}.
The design and performance of the BESIII detector are described in Refs.~\cite{BESIII:2009fln, BES:2001vqx, Yu:2016cof}.
The simulated Monte Carlo (MC) samples are produced with a {\sc geant4}-based software package~\cite{GEANT4:2002zbu}, which includes the geometric description of the BESIII detector and the
detector response. Furthermore, the simulation models the beam
energy spread and initial state radiation in the $e^+e^-$
annihilations with generator {\sc
kkmc}~\cite{Jadach:2000ir}. 

A sample of signal MC events for the process $\jpsi\to\XXb$ is used to determine detection efficiencies and optimize event selection criteria. The parameters of the angular distribution in the signal MC sample are obtained from the previous measurements~\cite{BESIII:2021ypr}.
To determine the normalization factors in the maximum likelihood fit, an exclusive MC sample is generated uniformly in phase space. 
To reduce the discrepancies between the experimental data and MC simulation and to minimize systematic uncertainties due to simulation limits, a control sample of $\jpsi\to\XXb$ where only one hyperon is reconstructed is used to estimate differences in hyperon reconstruction and event selection. 
This estimated discrepancy is taken into account by a weighting factor $f$, which is correlated to the angular distribution ($\cos\theta_\Xi$), of the events in the simulated sample, where $\theta_\Xi$ is the $\Xi^-$ scattering angle with respect to the $e^+$ beam in the $e^+e^-$ center-of-mass system.
An \textit{inclusive} MC sample is produced to estimate backgrounds. This sample includes the production of the $J/\psi$
resonance and the continuum processes incorporated in {\sc
kkmc}~\cite{Jadach:2000ir}. It also includes the subsequent $J/\psi$ particle decays, modeled with {\sc
evtgen}~\cite{Ping:2008zz} using branching fractions from the
Particle Data Group (PDG)~\cite{ParticleDataGroup:2024cfk}, when available
or otherwise estimated with {\sc lundcharm}~\cite{Chen:2000tv}.

The final event sample is obtained by a full reconstruction of the $\XXb\to\Lambda\pi^{-}\lambar\pi^{+}\to p\pi^{-}\pi^{-}\bar{p}\pi^{+}\pi^{+}$ process. 
For this purpose, at least three positively and three negatively charged tracks are required, where all tracks must fulfill the multilayer drift chamber (MDC) reconstruction requirement $|\cos\theta_{\rm LAB}| < 0.93$, where $\theta_{\rm LAB}$ is the polar angle relative to the positron beam direction. 
From this subset, the $\xim(\xip)$ candidates are reconstructed using two negative (positive) and one positive (negative) charged tracks. 
The two $\Xi$ decay chains are reconstructed separately, and are described here for the sequence $\xim\to\Lambda\pi^-\to p\pi^-\pi^-$.
To select the candidate $\xim$, one proton and two $\pi^-$ mesons must be identified. 
Events containing candidate protons with momenta larger than 0.3~GeV/$c$ or pions with momenta less than 0.3~GeV/$c$ are selected, which corresponds to nonoverlapping momentum distributions of protons and pions of the signal process.
The $\Lambda$ and $\xim$ particles are reconstructed from vertex fits by first combining the $p\pi^-_i$ pair to a $\Lambda$ and then the $\Lambda\pi^-_j(i\neq j)$ pair to a $\Xi^-$ candidate. The fits take into account the nonzero flight paths of the hyperons, which can give rise to different production and decay positions.
The remaining candidates are then required to pass additional kinematic constraints to reject non-$\Lambda$, non-$\Xi$, and combinatorial background contributions with $\delta<0.016$ GeV/$c^2$, where
\begin{equation}
\begin{array}{cc}
    \label{eq:delta}
    \delta &= \sqrt{(m_{\Lambda\pi^-} - m_{\Xi^-})^2 + {R}^2(m_{p\pi^-} - m_{\Lambda})^2}. \\
    \end{array}
\end{equation}
\noindent Here the $m_{\xim}$ and $m_\Lambda$ are the PDG masses ~\cite{ParticleDataGroup:2024cfk} and $m_{\Lambda\pi^-}(m_{p\pi^-})$ is the invariant mass of the $\Lambda\pi^-(p\pi^-)$ combination.
The
$R = \sigma_{\Xi^-}/\sigma_{\Lambda}$ is the ratio of the resolution of the $m_{\Lambda\pi^-}$ distribution over that of the $m_{p\pi^-}$ distribution, where the resolution is extracted by fitting the mass distribution with a sum of two Gaussian functions.
The selection criterion for $\delta$ is validated using a figure of merit (FOM) defined as $S/\sqrt{S+B}$, where $S$ represents the number of signal events estimated from the MC simulation, and $B$ denotes the background contribution. The quantity $S+B$ is derived from data events.
Events where the reconstruction algorithm gives a negative decay length reading are rejected. 
A four-constraint (4C) kinematic fit requiring energy and momentum conservation is imposed on the $\EE\to\jpsi\to\XXb$ system, and only events satisfying the optimized requirement of $\chifc<200$ by FOM are retained for further analysis. This kinematic constraint effectively removes background processes with the same charged final-state topology but containing extra neutral particles, and improves hyperon mass resolution. 

After applying all aforementioned selection criteria, $5.8\times10^5\ \EE\to\XXb$ candidates remain in the final sample. 
The amount of background contamination from this sample is evaluated with a two-dimensional sideband using a looser requirement of $\delta(\overline\delta)$. 
Two regions in the distribution $m_{\Lambda\pi^-}$ versus $m_{\bar\Lambda\pi^+}$ are selected as the estimation areas for the sidebands where the lower and upper limitations correspond to $1.274<m_{\Lambda\pi^-(\bar\Lambda\pi^+)}<1.306~{\rm GeV/}c^2$ and $1.338<m_{\Lambda\pi^-(\bar\Lambda\pi^+)}<1.370~{\rm GeV/}c^2$, respectively. 
The number of sideband background events is found to be $2,034\pm45$, or 0.4\%, after normalizing the data events to the range of $\delta(\overline\delta)<0.016$ GeV/$c^2$.

\begin{figure}[!htbp]
    \centering
    \includegraphics[width=0.49\linewidth]{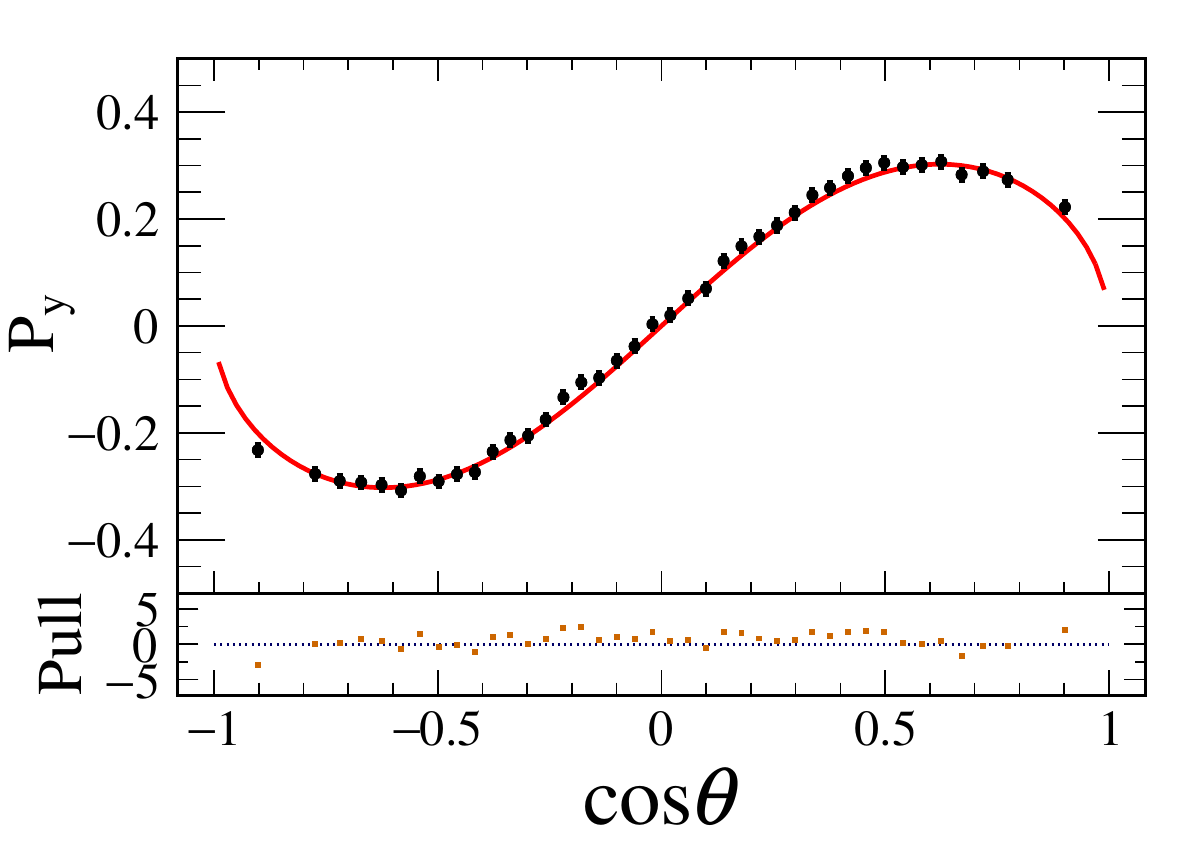}
    \includegraphics[width=0.49\linewidth]{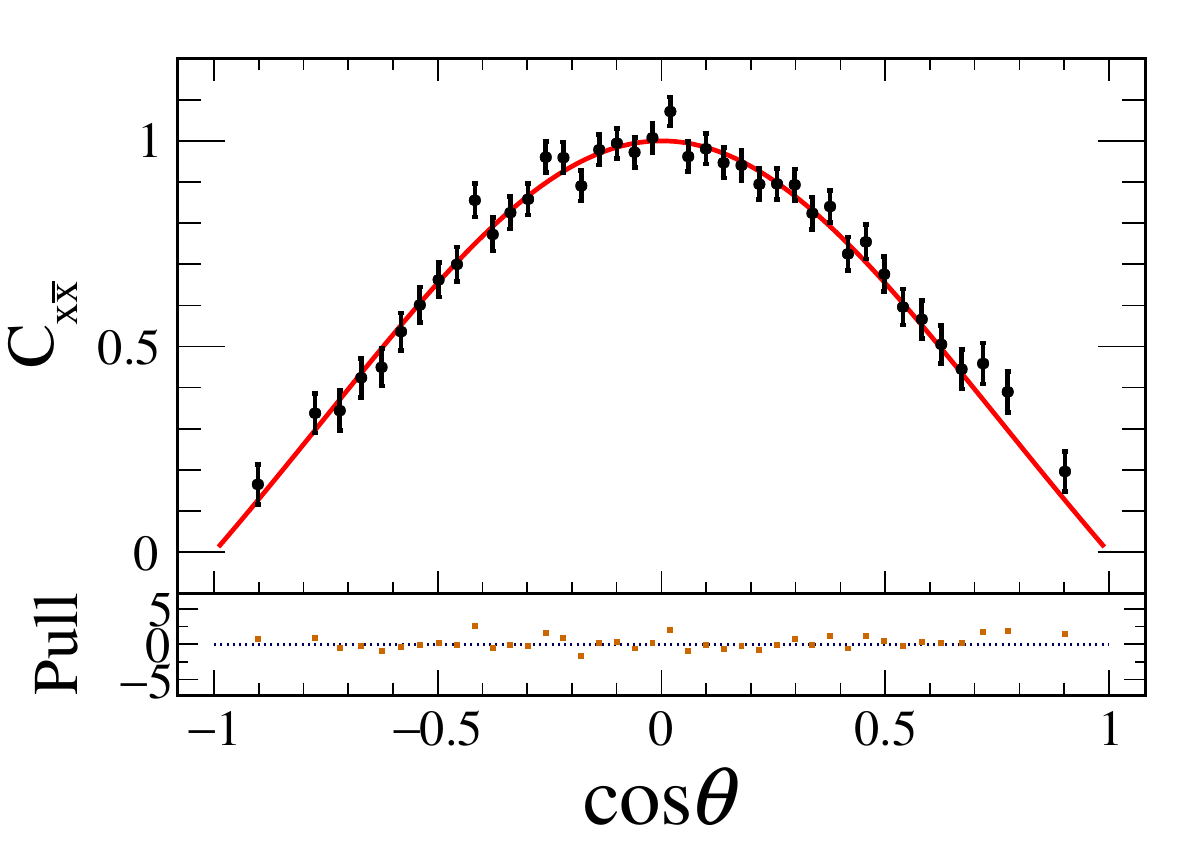}
    \includegraphics[width=0.49\linewidth]{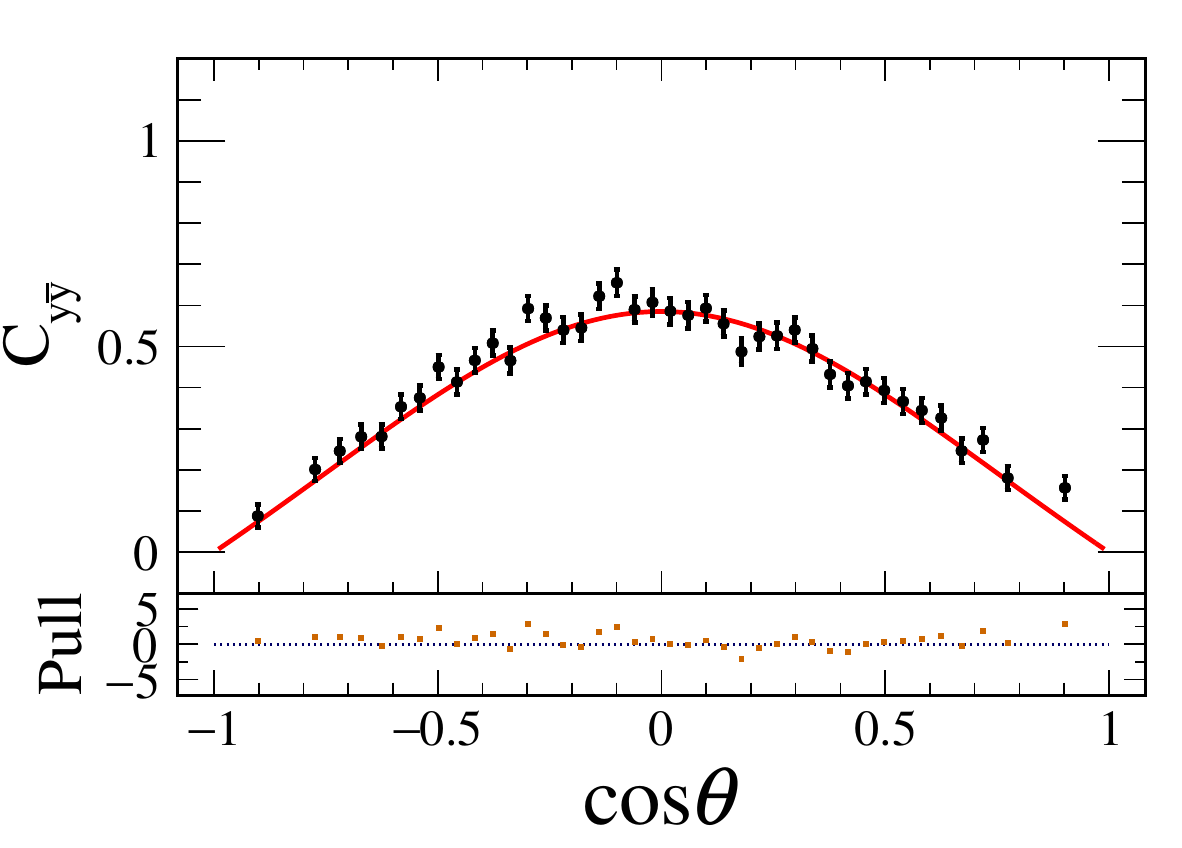}
    \includegraphics[width=0.49\linewidth]{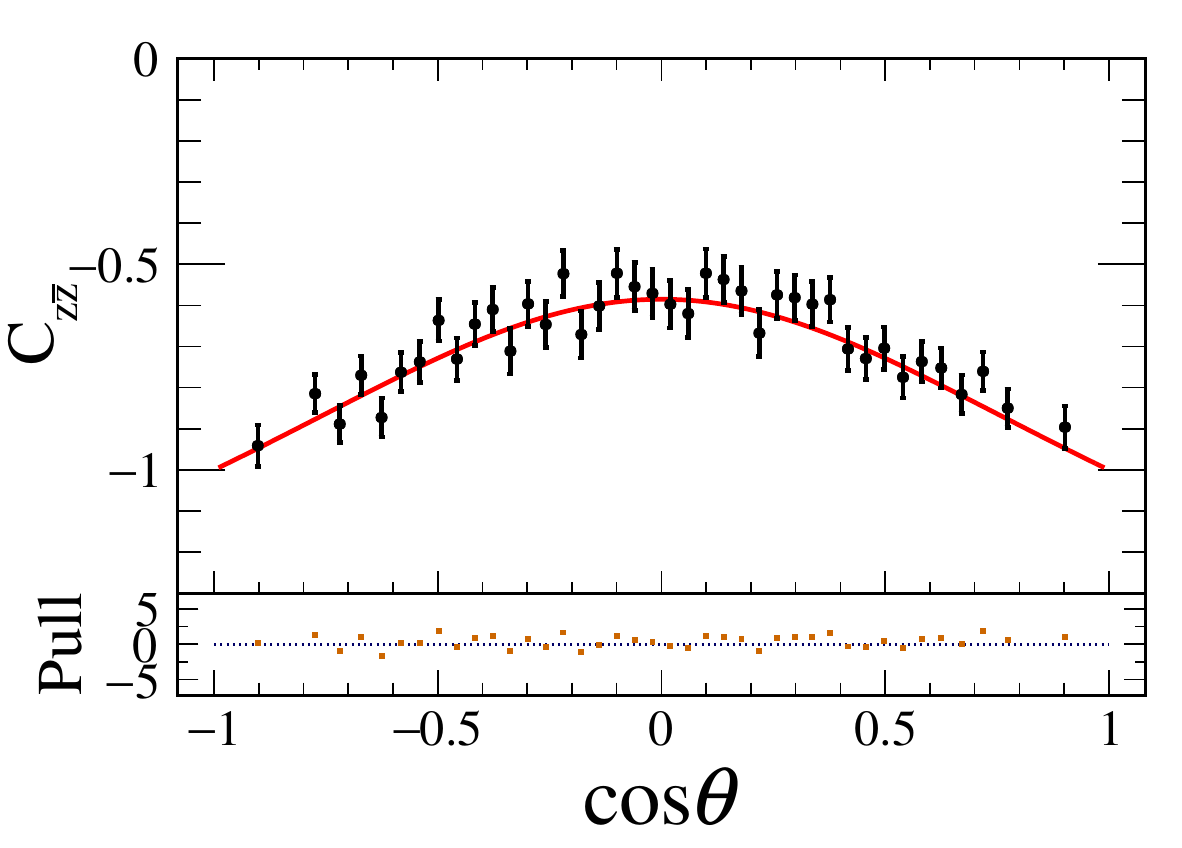}
    \caption{ 
    Acceptance distributions of transverse polarization and spin 
correlations as a function of $\cos\theta_\Xi$. The black points with error bars show the results from local fits in bins of $\cos\theta_\Xi$, 
where the error bars represent statistical uncertainties. The red curves correspond to the global fit using the 
production parameters $\alpha_\psi$ and $\Delta \Phi$.}
    \label{fig:fit}
\end{figure}

The formalism developed to determine the joint angular distribution~\cite{Perotti:2018wxm, BESIII:2021ypr} allows for a direct measurement of two $\jpsi\to\XXb$ production parameters, $\alpha_\psi$ and $\Delta\Phi$, and six weak decay parameters ($\alpha_{\Lambda},\ \alpha_{\Xi},\ \phi_\Xi$) of the $\Xi^-\to\Lambda\pi^-$, $\Lambda\to p\pi^-$ and their charge conjugated parameters ($\overline\alpha_{\Lambda},\ \overline\alpha_{\Xi},\ \overline\phi_\Xi$). 
These eight parameters are fully determined from the nine helicity angles $\boldsymbol{\xi}=\{\theta_\Xi,\, \theta_\Lambda,\, \varphi_\Lambda,\, \theta_{\bar\Lambda},\, \varphi_{\bar\Lambda},\, \theta_p,\, \varphi_p,\, \theta_{\bar p},\, \varphi_{\bar p}\}$, where except for $\theta_\Xi$, 
 the other eight parameters are the polar and azimuthal helicity angles of the $\Lambda$, $\lambar$, proton, and anti-proton, respectively~\cite{Perotti:2018wxm, BESIII:2021ypr}.
A maximum log-likelihood fit, following Refs.~\cite{BESIII:2021ypr, BESIII:2022lsz}, is used, where the sideband background events are subtracted from the fit. 
The MC samples are approximately thirty times larger than data to ensure that uncertainties due to MC statistics are sufficiently minimized.
To illustrate the fit quality, the transverse polarization and the diagonal spin correlations defined in Ref.~\cite{Perotti:2018wxm} are shown in Fig.~\ref{fig:fit}.
The fit results are summarized in Table~\ref{table:sumres}.

\begin{table*}[hbtp]
\centering
\caption{
Results 
for the $J/\psi\to\Xi^{-}\bar{\Xi}^+$ fitted parameters and $CP$ observables,
as well as the comparisons with the previous best measurements. The first and second uncertainties are statistical and systematic, respectively.}
\begin{tabular*}{\textwidth}{@{\extracolsep{\fill}}lcc}
\hline\hline
Parameter &\multicolumn{1}{c}{This work} & \multicolumn{1}{c}{Previous best result}  \\
\hline
$\alpha_{\psi}$ &               $\phantom{-}0.5851\pm0.0044\pm0.0034$ &    $\phantom{-}0.611 \pm 0.007 ^{+0.013}_{-0.007}$~\cite{BESIII:2023jhj}  \\
$\Delta\Phi$~(rad) &            $\phantom{-}1.2205\pm0.0159\pm0.0056$ &    $\phantom{-}1.30 \pm 0.03 ^{+0.02}_{-0.03}$~\cite{BESIII:2023jhj} \\
$\alpha_{\Xi}$ &                $-0.3813\pm0.0026\pm0.0005$           &    $-0.367 \pm 0.004 ^{+0.003}_{-0.004}$~\cite{BESIII:2023jhj}  \\
$\overline{\alpha}_{\Xi}$ &     $\phantom{-}0.3873\pm0.0026\pm0.0006$ &    $\phantom{-}0.374 \pm 0.004 ^{+0.003}_{-0.004}$~\cite{BESIII:2023jhj} \\
$\phi_{\Xi}$~(rad) &            $-0.0008\pm0.0072\pm0.0010$           &    $-0.016 \pm 0.012 ^{+0.004}_{-0.008}$~\cite{BESIII:2023jhj} \\
$\overline{\phi}_{\Xi}$~(rad) & $\phantom{-}0.0020\pm0.0072\pm0.0006$ &    $\phantom{-}0.010 \pm 0.012 ^{+0.003}_{-0.013}$~\cite{BESIII:2023jhj}  \\
$\alpha_{\Lambda}$ &            $\phantom{-}0.7434\pm0.0039\pm0.0015$ &    $\phantom{-}0.7519 \pm 0.0036\pm 0.0024$\hfill \cite{BESIII:2022qax}  \\
$\overline{\alpha}_{\Lambda}$ & $-0.7478\pm0.0038\pm0.0015$           &    $-0.7559 \pm 0.0036\pm 0.0030$\hfill \cite{BESIII:2022qax}  \\
\hline
$(\xi_P - \xi_S)\times10^{-2}$~(rad)               & $-0.2\pm1.2\pm0.1$             & $\phantom{-}0.7\pm2.0^{+1.8}_{-0.5}$~\cite{BESIII:2023jhj}  \\
$(\delta_P - \delta_S)\times10^{-2}$~(rad)         & $\phantom{-}0.3\pm1.2\pm0.2$   & $\phantom{-}3.3\pm2.0 ^{+0.8}_{-1.2}$~\cite{BESIII:2023jhj} \\
\hline
$A_{CP}^{\Xi}\times10^{-3}$                  & $-7.8\pm4.8\pm0.8$ & $-9\pm8^{+7}_{-2}$~\cite{BESIII:2023jhj} \\
$\Delta\phi_{CP}^{\Xi}\times10^{-3}$~(rad)   & $\phantom{-}0.6\pm5.1\pm0.2$ & $-3\pm8^{+3}_{-7}$~\cite{BESIII:2023jhj} \\
$A_{CP}^{\Lambda}\times10^{-3}$              & $-2.9\pm4.3\pm0.7$ & $-2.5\pm4.6\pm1.2$~\cite{BESIII:2022qax} \\
\hline
$\langle\alpha_\Xi\rangle$         & $-0.3843\pm0.0018\pm0.0005$ & $-0.373\pm0.005\pm0.002$~\cite{BESIII:2021ypr} \\
$\langle\phi_\Xi\rangle$~(rad)     & $-0.0014\pm0.0050\pm0.0008$ & $\phantom{-}0.016\pm0.014\pm0.007$~\cite{BESIII:2021ypr} \\
$\langle\alpha_\Lambda\rangle$     & $\phantom{-}0.7456\pm0.0022\pm0.0014$ & $\phantom{-}0.7542\pm0.0010\pm0.0024$~\cite{BESIII:2022qax} \\
\hline\hline
\end{tabular*}
\label{table:sumres}
\end{table*}

 \begin{table}[htbp]
     \caption{Absolute values of systematic uncertainties on the measurements of the fitted parameters in the $\jpsi\to\XXb$ process. All entries have been multiplied by $10^3$.}
     \centering
     \label{tab:sys}
     \begin{tabular}{ccccc}
 	\hline \hline
    Parameter	         &Fit method & Reconstruction &	Background & Total \\ \hline
$\alpha_{\psi}$	             &	0.2	&    3.0	 &1.6	   &  3.4\\ \hline
$\Delta{\Phi}$~(rad)	             &	0.0	&    4.6	 &3.3	   &  5.6\\ \hline
$\alpha_{\Xi}$	             &	0.3	&    0.2	 &0.4	   &  0.5\\ \hline
$\overline\alpha_{\Xi}$	     &	0.3	&    0.2	 &0.5	   &  0.6\\ \hline
$\phi_{\Xi}$~(rad)	             &	0.9	&    0.1	 &0.3	   &  1.0\\ \hline
$\overline\phi_{\Xi}$~(rad)	     &	0.5	&    0.1	 &0.3	   &  0.6\\ \hline
$\alpha_{\Lambda}$	         &	0.5	&    0.3	 &1.3	   &  1.5\\ \hline
$\overline\alpha_{\Lambda}$	 &	0.5	&    0.3	 &1.4	   &  1.5\\
 	\hline \hline
    \end{tabular}
    \end{table}

The event sample is eight times larger than the previous study in Ref.~\cite{BESIII:2021ypr}, which improves the statistical uncertainty by more than a factor of 2.5. 
The systematic uncertainties can be divided into two sources. One is from the discrepancy between data and MC simulation, including uncertainties on the $\Xi$ reconstruction and background estimations. The other source comes from the potential bias of the fit method.
The systematic uncertainty due to the $\XXb$ reconstruction includes contributions from track selection, $\Lambda$ reconstruction, the 4C kinematic fit, $\Xi$ mass window selection and $\Xi$ decay length requirement. It is evaluated using a control sample of $\jpsi\to\XXb$, where only one hyperon is reconstructed, and depends on the polar angles $\theta$ of $\Xi^-$ and $\xip$. By applying Gaussian sampling to the weighting factor referred to as the above $f$, the maximum likelihood fit is repeated one hundred times. The standard deviation of the fit results, extracted from a Gaussian fit, is taken as the systematic uncertainty. 
The background estimation uncertainty is assessed by including and excluding the sideband background events. The difference between the two results is assigned as a systematic uncertainty.
The fit bias is examined using three hundred pseudoexperiments, generated with input parameters identical to those reported in Ref.~\cite{BESIII:2021ypr}. 
The systematic uncertainty from the fit procedure is taken to be the absolute difference between the input and output values in the pseudoexperiments.
All systematic uncertainties are listed in Table~\ref{tab:sys}. The total systematic uncertainty is obtained by summing the individual contributions in quadrature. 


The $\XXb$ production and decay parameters of $\alpha_\psi,\ \Delta\Phi,\ \alpha_\Xi,\ \overline\alpha_\Xi$, $\phi_\Xi$, and $\overline\phi_\Xi$ are determined with the most accurate statistical and systematic uncertainties available, as shown in Table~\ref{table:sumres} and the measured values of these decay parameters are in agreement with previous results. 

\begin{figure}[htbp]
    \centering
\includegraphics[width=1\linewidth]{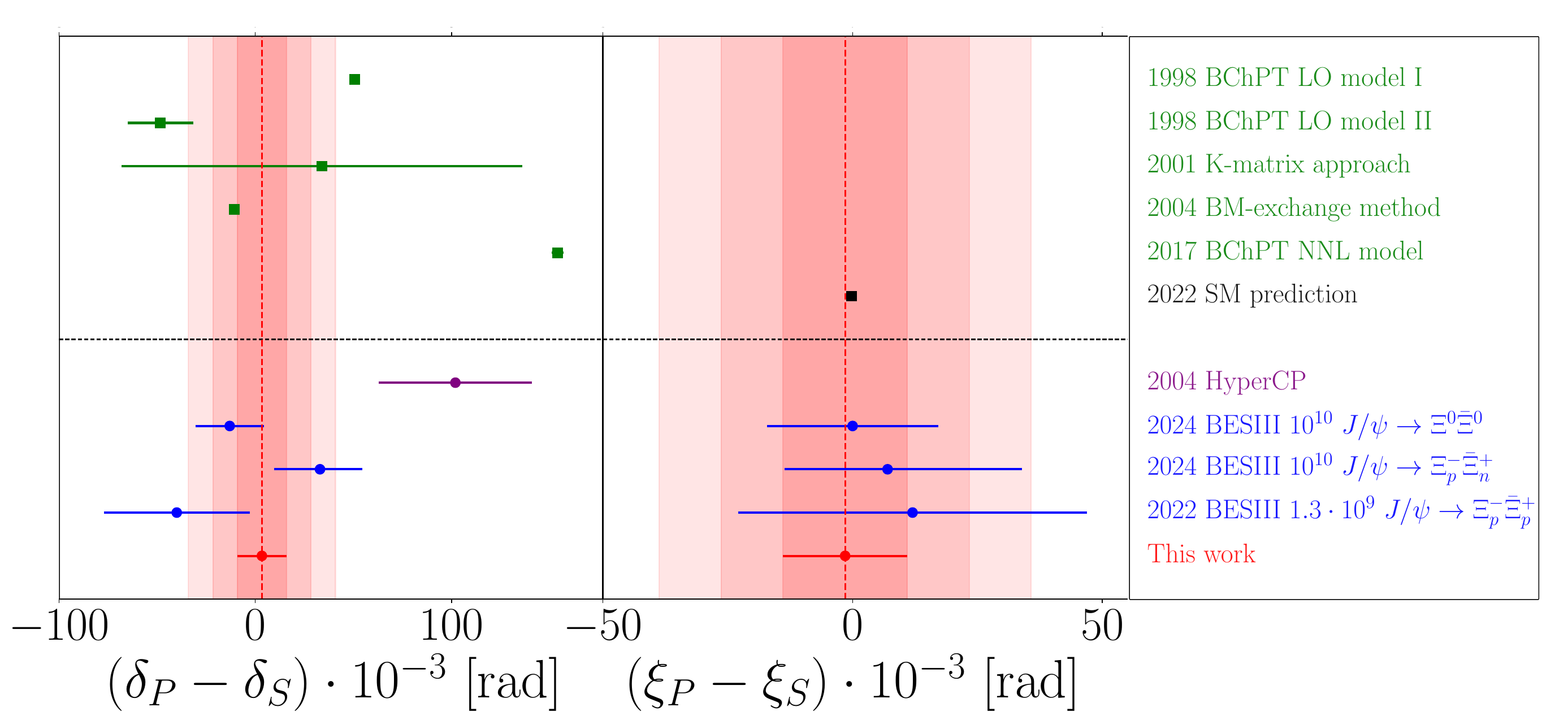}
    \caption{Strong and weak phase differences in hyperon processes.
    The red dots with error bars represent the results in this work, which indicate the best results of $(\delta_P-\delta_S)_{\xim}$ and $(\xi_P-\xi_S)_{\xim}$. The blue dots with error bars denote the previous BESIII results from the processes $\jpsi\to\XXb\to\Lambda(p\pi^-)\pi^-\bar{\Lambda}(\bar{n}\pi^0)\pi^+$~\cite{BESIII:2023jhj},  $\jpsi\to\XXb\to\Lambda(p\pi^-)\pi^-\bar{\Lambda}(\bar{p}\pi^+)\pi^+$~\cite{BESIII:2021ypr}, and $J/\psi\to\Xi^0\bar{\Xi}^0$~\cite{Besiii:2023drj}, respectively. The purple dot with error bar indicates the result of $(\delta_P-\delta_S)_{\xim}$ measured by the HyperCP Collaboration~\cite{HyperCP:2004zvh}. 
    The black square with error bar implies the predictions from the SM for weak phase difference~\cite{Salone:2022lpt}.
    The green squares with error bars denote the theoretical predictions on the strong phase differences using various models~\cite{Kamal:1998se, Datta:1998pv, Tandean:2000dx, Barros:2004pw, Huang:2017bmx}.
    The inner, middle, and outer bands represent 68.2\%, 95.4\%, and 99.7\% confidence intervals of variables in this work, respectively.}
    \label{fig:acp}
\end{figure}

The measurement of the average values of $\langle\alpha_\Xi\rangle$ and $\langle\phi_\Xi\rangle$ enables a direct determination of the strong and weak phase differences via Eqs.~(\ref{eq:strP}) and~(\ref{eq:weakP}). Our result on the strong phase difference, $(\delta_P - \delta_S) = (0.3\pm1.2\pm0.2)\times10^{-2}$~rad, is consistent with zero within the uncertainties. It is in agreement with the earlier BESIII results, but is different from the HyperCP value by 2.3$\sigma$~\cite{HyperCP:2004zvh}, and disfavors the latest theoretical value by 11.8$\sigma$~\cite{Huang:2017bmx}.
The weak phase difference is determined to be $(\xi_P-\xi_S)=(-0.2\pm1.2\pm0.1)\times10^{-2}$~rad. To our knowledge, this is the most precise weak phase determination for any baryon weak decay, but its precision is still 2 orders of magnitude lower than the corresponding SM prediction $(\xi_P-\xi_S)_{\rm SM}=-(2.1\pm1.7)\times10^{-4}$~rad~\cite{Salone:2022lpt}.
Our results of $(\delta_P-\delta_S)$ and $(\xi_P-\xi_S)$, and the theoretical predictions and the other measurements are shown in Fig.~\ref{fig:acp}.

The extracted $\XXb$ decay parameters allow for two $CP$ symmetry tests. 
The asymmetry $A_{CP}^{\Xi}$ is measured to be $(-7.8\pm4.8\pm0.8)\times10^{-3}$. The corresponding SM prediction is $0.5\times10^{-5} < A_{CP,{\rm SM}}^{\Xi} < 6\times10^{-5}$~\cite{Salone:2022lpt}. Our result is consistent with the SM prediction within 1.5$\sigma$. 
Another $CP$ test, defined by the parameters $\phi_\Xi$ and $\overline\phi_\Xi$, yields $\Delta\phi_{CP}^{\Xi} =(0.6\pm5.1\pm0.2)\times10^{-3}~\text{rad}$. The first and second uncertainties are statistical and systematic, respectively.

The $\LLb$ decay parameters $\alpha_\Lambda$ and $\overline\alpha_\Lambda$ are also measured in the $\jpsi\to\XXb\to\Lambda\overline\Lambda\pi^-\pi^+$ decay. 
Despite being based on a data sample six times smaller, our results are statistically comparable with the results obtained from $\jpsi\to\LLb$~\cite{BESIII:2022qax}, 
while exhibiting significantly reduced systematic uncertainties.
This is explained by the product with $\alpha_\Lambda$ and spin polarization.
The correlation coefficient between the parameters is $\rho(\alpha_\Lambda,\ \overline\alpha_\Lambda)_{\jpsi\to\XXb}=0.372$, significantly lower than $\rho(\alpha_\Lambda,\ \overline\alpha_\Lambda)_{\jpsi\to\LLb}=0.850$. This implies that our measurement yields a more precise determination of $A_{CP}$, but a less precise estimate of the average value of $\alpha_\Lambda$. 
The $CP$ asymmetry of the $\Lambda$ hyperon is measured to be $A_{CP}^{\Lambda} = (-2.9\pm4.3\pm0.7)\times10^{-3}$ representing the most precise determination to date.
The $CP$ violation test reveals no significant deviation from $CP$ symmetry.
The average value is determined to be $\langle\alpha_\Lambda\rangle = (\alpha_\Lambda-\overline\alpha_\Lambda)/2 = 0.7456\pm0.0022\pm0.0014$, indicating a discrepancy of 2.3$\sigma$ compared to the result from $\jpsi\to\LLb$.

In summary, by using $(10\,087\pm44)\times10^{6}$ $\jpsi$ events~\cite{BESIII:2021cxx}, 
the most precise measurements of the $\Xi$ strong phase and weak phase differences are determined in the $\jpsi\to\XXb$ decay, as summarized in Table~\ref{table:sumres}. 
Our measurements are consistent with the previous BESIII results of the double-strange hyperon~\cite{BESIII:2021ypr, BESIII:2022lsz, BESIII:2023jhj}, with significantly improved accuracy.

Nonetheless, our results of the  strong and weak phase differences in baryon weak
decays remain consistent with zero.
Probing possible nonzero values would require substantially larger data samples; future facilities such as the next-generation tau-charm physics facility~\cite{Charm-TauFactory:2013cnj,Achasov:2023gey,Salone:2022lpt} and PANDA~\cite{PANDA:2020zwv} will provide important steps in this direction.
The investigation of potential systematic sources put forth by this analysis will be useful for the super tau-charm facility in particular~\cite{Achasov:2023gey}, which aims to produce a sample of $10^{12}$ $J/\psi$. 

\textit{Acknowledgments}—Thanks to Jusak Tandean for useful discussions about the theoretical approaches related to the strong phase difference. 
The BESIII Collaboration thanks the staff of BEPCII~\cite{BEPC:2025xni} and the IHEP computing center for their strong support. This work is supported in part by National Key R\&D Program of China under Contracts No. 2023YFA1606000, No. 2023YFA1606704; National Natural Science Foundation of China (NSFC) under Contracts No. 12247101, No. 11635010, No. 11935015, No. 11935016, No. 11935018, No. 12025502, No. 12035009, No. 12035013, No. 12061131003, No. 12192260, No. 12192261, No. 12192262, No. 12192263, No. 12192264, No. 12192265, No. 12221005, No. 12225509, No. 12235017, No. 12361141819; 
the Fundamental Research Funds for the Central Universities No. lzujbky-2025-ytA05, No. lzujbky-2025-it06, No. lzujbky-2024-jdzx06; the Natural Science Foundation of Gansu Province No. 22JR5RA389, No. 25JRRA799; the "111 Center" under Grant No. B20063; 
Beijing Natural Science Foundation of China (BNSF) under Contract No. IS23014; the Chinese Academy of Sciences (CAS) Large-Scale Scientific Facility Program; the Strategic Priority Research Program of Chinese Academy of Sciences under Contract No. XDA0480600; CAS under Contract No. YSBR-101; 100 Talents Program of CAS; The Institute of Nuclear and Particle Physics (INPAC) and Shanghai Key Laboratory for Particle Physics and Cosmology; ERC under Contract No. 758462; German Research Foundation DFG under Contract No. FOR5327; Istituto Nazionale di Fisica Nucleare, Italy; Knut and Alice Wallenberg Foundation under Contracts No. 2021.0174, No. 2021.0299; Ministry of Development of Turkey under Contract No. DPT2006K-120470; National Research Foundation of Korea under Contract No. NRF-2022R1A2C1092335; National Science and Technology fund of Mongolia; Polish National Science Centre under Contract No. 2024/53/B/ST2/00975; STFC (United Kingdom); Swedish Research Council under Contract No. 2019.04595; the Olle Engkvist Foundation under Contract No. 200-0605; U. S. Department of Energy under Contract No. DE-FG02-05ER41374.

\textit{Data availability}—The data that support the findings of this article are not publicly available upon publication
because  it is not technically feasible and/or the cost of preparing, depositing,  and hosting the data would  be prohibitive
within  the terms  of this research project. The data are available from the authors upon reasonable request.

\input{apstemplate.bbl}

\section*{End Matter}
This End Matter introduces the correlation coefficients of the fit parameters, as shown in Table~\ref{table:maincorrelations}, which is helpful for understanding the correlations among the parameters.

\begin{table}[htbp]
\caption{Correlation coefficients of the fit results.}
\label{table:maincorrelations}
\begin{tabular}{|c|c| c | c | c | c | c | c | c | c |}
\hline\hline
                            & $\alpha_{\psi}$ & $\Delta\Phi$ & $\alpha_{\Xi}$ & $\overline\alpha_{\Xi}$ & $\phi_\Xi$ & $\overline{\phi}_\Xi$ & $\alpha_{\Lambda}$ & $\overline\alpha_{\Lambda}$ \\  \hline
$\alpha_{\psi}$             & 1.000      & 0.394 & $-0.023$  & 0.016      & $-0.016$     & 0.017      & $-0.116$        & 0.107         \\ 
$\Delta\Phi$                &            & 1.000 & $-0.035$  & 0.021      & 0.010      & $-0.002$     & $-0.148$        & 0.135         \\ 
$\alpha_{\Xi}$              &            &       & 1.000   & 0.035      & $-0.002$     & 0.001      & 0.312         & 0.078         \\ 
$\overline\alpha_{\Xi}$     &            &       &         & 1.000      & 0.000      & $-0.001$     & 0.083         & 0.316         \\ 
$\phi_\Xi$                  &            &       &         &            & 1.000      & 0.029      & 0.004         & 0.001         \\ 
$\overline{\phi}_\Xi$       &            &       &         &            &            & 1.000      & 0.001         & 0.006         \\ 
$\alpha_{\Lambda}$          &            &       &         &            &            &            & 1.000         & 0.372         \\ 
$\overline\alpha_{\Lambda}$ &            &       &         &            &            &            &               & 1.000         \\
\hline\hline
\end{tabular}
\end{table}

\begin{widetext}
\begin{center}
\small
M.~Ablikim$^{1}$\BESIIIorcid{0000-0002-3935-619X},
M.~N.~Achasov$^{4,b}$\BESIIIorcid{0000-0002-9400-8622},
P.~Adlarson$^{81}$\BESIIIorcid{0000-0001-6280-3851},
X.~C.~Ai$^{86}$\BESIIIorcid{0000-0003-3856-2415},
R.~Aliberti$^{39}$\BESIIIorcid{0000-0003-3500-4012},
A.~Amoroso$^{80A,80C}$\BESIIIorcid{0000-0002-3095-8610},
Q.~An$^{77,64,\dagger}$,
Y.~Bai$^{62}$\BESIIIorcid{0000-0001-6593-5665},
O.~Bakina$^{40}$\BESIIIorcid{0009-0005-0719-7461},
Y.~Ban$^{50,g}$\BESIIIorcid{0000-0002-1912-0374}
% ...
\\
\vspace{0.2cm}
(BESIII Collaboration)\\
\vspace{0.2cm} {\it
$^{1}$ Institute of High Energy Physics, Beijing 100049, People's Republic of China\\
% ...
$^{4}$ Budker Institute of Nuclear Physics SB RAS (BINP), Novosibirsk 630090, Russia\\
% ...
$^{39}$ Johannes Gutenberg University of Mainz, Johann-Joachim-Becher-Weg 45, D-55099 Mainz, Germany\\
$^{40}$ Joint Institute for Nuclear Research, 141980 Dubna, Moscow region, Russia\\
% ...
$^{50}$ Peking University, Beijing 100871, People's Republic of China\\
% ...
$^{62}$ Southeast University, Nanjing 211100, People's Republic of China\\
% ...
$^{64}$ State Key Laboratory of Particle Detection and Electronics, Beijing 100049, Hefei 230026, People's Republic of China\\
% ...
$^{77}$ University of Science and Technology of China, Hefei 230026, People's Republic of China\\
% ...
$^{80}$ University of Turin and INFN, (a)University of Turin, I-10125, Turin, Italy; (b)University of Eastern Piedmont, I-15121, Alessandria, Italy; (v)INFN, I-10125, Turin, Italy\\
$^{81}$ Uppsala University, Box 516, SE-75120 Uppsala, Sweden\\
% ...
$^{86}$ Zhengzhou University, Zhengzhou 450001, People's Republic of China\\
\vspace{0.2cm}
$^{\dagger}$ Deceased.\\
% ...
$^{b}$ Also at the Novosibirsk State University, Novosibirsk, 630090, Russia.\\
% ...
$^{g}$ Also at State Key Laboratory of Nuclear Physics and Technology, Peking University, Beijing 100871, People's Republic of China.\\
% ...
}
\end{center}
\end{widetext}

\end{document}

%% file: apstemplate.bbl
%

%% file: apstemplate.bbl
\begin{thebibliography}{49}%
\makeatletter
\providecommand \@ifxundefined [1]{%
 \@ifx{#1\undefined}
}%
\providecommand \@ifnum [1]{%
 \ifnum #1\expandafter \@firstoftwo
 \else \expandafter \@secondoftwo
 \fi
}%
\providecommand \@ifx [1]{%
 \ifx #1\expandafter \@firstoftwo
 \else \expandafter \@secondoftwo
 \fi
}%
\providecommand \natexlab [1]{#1}%
\providecommand \enquote  [1]{``#1''}%
\providecommand \bibnamefont  [1]{#1}%
\providecommand \bibfnamefont [1]{#1}%
\providecommand \citenamefont [1]{#1}%
\providecommand \href@noop [0]{\@secondoftwo}%
\providecommand \href [0]{\begingroup \@sanitize@url \@href}%
\providecommand \@href[1]{\@@startlink{#1}\@@href}%
\providecommand \@@href[1]{\endgroup#1\@@endlink}%
\providecommand \@sanitize@url [0]{\catcode `\\12\catcode `\$12\catcode `\&12\catcode `\#12\catcode `\^12\catcode `\_12\catcode `\%12\relax}%
\providecommand \@@startlink[1]{}%
\providecommand \@@endlink[0]{}%
\providecommand \url  [0]{\begingroup\@sanitize@url \@url }%
\providecommand \@url [1]{\endgroup\@href {#1}{\urlprefix }}%
\providecommand \urlprefix  [0]{URL }%
\providecommand \Eprint [0]{\href }%
\providecommand \doibase [0]{https://doi.org/}%
\providecommand \selectlanguage [0]{\@gobble}%
\providecommand \bibinfo  [0]{\@secondoftwo}%
\providecommand \bibfield  [0]{\@secondoftwo}%
\providecommand \translation [1]{[#1]}%
\providecommand \BibitemOpen [0]{}%
\providecommand \bibitemStop [0]{}%
\providecommand \bibitemNoStop [0]{.\EOS\space}%
\providecommand \EOS [0]{\spacefactor3000\relax}%
\providecommand \BibitemShut  [1]{\csname bibitem#1\endcsname}%
\let\auto@bib@innerbib\@empty
\bibitem [{\citenamefont {Sakharov}(1967)}]{Sakharov:1967dj}%
  \BibitemOpen
  \bibfield  {author} {\bibinfo {author} {\bibfnamefont {A.~D.}\ \bibnamefont {Sakharov}},\ }\href {https://doi.org/10.1070/PU1991v034n05ABEH002497} {\bibfield  {journal} {\bibinfo  {journal} {Pis'ma Zh. Eksp. Teor. Fiz.}\ }\textbf {\bibinfo {volume} {5}},\ \bibinfo {pages} {32} (\bibinfo {year} {1967})}\BibitemShut {NoStop}%
\bibitem [{\citenamefont {Cabibbo}(1963)}]{Cabibbo:1963yz}%
  \BibitemOpen
  \bibfield  {author} {\bibinfo {author} {\bibfnamefont {N.}~\bibnamefont {Cabibbo}},\ }\href {https://doi.org/10.1103/PhysRevLett.10.531} {\bibfield  {journal} {\bibinfo  {journal} {Phys. Rev. Lett.}\ }\textbf {\bibinfo {volume} {10}},\ \bibinfo {pages} {531} (\bibinfo {year} {1963})}\BibitemShut {NoStop}%
\bibitem [{\citenamefont {Kobayashi}\ and\ \citenamefont {Maskawa}(1973)}]{Kobayashi:1973fv}%
  \BibitemOpen
  \bibfield  {author} {\bibinfo {author} {\bibfnamefont {M.}~\bibnamefont {Kobayashi}}\ and\ \bibinfo {author} {\bibfnamefont {T.}~\bibnamefont {Maskawa}},\ }\href {https://doi.org/10.1143/PTP.49.652} {\bibfield  {journal} {\bibinfo  {journal} {Prog. Theor. Phys.}\ }\textbf {\bibinfo {volume} {49}},\ \bibinfo {pages} {652} (\bibinfo {year} {1973})}\BibitemShut {NoStop}%
\bibitem [{\citenamefont {Bernreuther}(2002)}]{Bernreuther:2002uj}%
  \BibitemOpen
  \bibfield  {author} {\bibinfo {author} {\bibfnamefont {W.}~\bibnamefont {Bernreuther}},\ }\href {https://doi.org/https://doi.org/10.1007/3-540-47895-7_7} {\bibfield  {journal} {\bibinfo  {journal} {Lect. Notes Phys.}\ }\textbf {\bibinfo {volume} {591}},\ \bibinfo {pages} {237} (\bibinfo {year} {2002})}\BibitemShut {NoStop}%
\bibitem [{\citenamefont {Canetti}\ \emph {et~al.}(2012)\citenamefont {Canetti}, \citenamefont {Drewes},\ and\ \citenamefont {Shaposhnikov}}]{Canetti:2012zc}%
  \BibitemOpen
  \bibfield  {author} {\bibinfo {author} {\bibfnamefont {L.}~\bibnamefont {Canetti}}, \bibinfo {author} {\bibfnamefont {M.}~\bibnamefont {Drewes}},\ and\ \bibinfo {author} {\bibfnamefont {M.}~\bibnamefont {Shaposhnikov}},\ }\href {https://doi.org/10.1088/1367-2630/14/9/095012} {\bibfield  {journal} {\bibinfo  {journal} {New J. Phys.}\ }\textbf {\bibinfo {volume} {14}},\ \bibinfo {pages} {095012} (\bibinfo {year} {2012})}\BibitemShut {NoStop}%
\bibitem [{\citenamefont {Christenson}\ \emph {et~al.}(1964)\citenamefont {Christenson}, \citenamefont {Cronin}, \citenamefont {Fitch},\ and\ \citenamefont {Turlay}}]{Christenson:1964fg}%
  \BibitemOpen
  \bibfield  {author} {\bibinfo {author} {\bibfnamefont {J.~H.}\ \bibnamefont {Christenson}}, \bibinfo {author} {\bibfnamefont {J.~W.}\ \bibnamefont {Cronin}}, \bibinfo {author} {\bibfnamefont {V.~L.}\ \bibnamefont {Fitch}},\ and\ \bibinfo {author} {\bibfnamefont {R.}~\bibnamefont {Turlay}},\ }\href {https://doi.org/10.1103/PhysRevLett.13.138} {\bibfield  {journal} {\bibinfo  {journal} {Phys. Rev. Lett.}\ }\textbf {\bibinfo {volume} {13}},\ \bibinfo {pages} {138} (\bibinfo {year} {1964})}\BibitemShut {NoStop}%
\bibitem [{\citenamefont {Aubert}\ \emph {et~al.}(2001)\citenamefont {Aubert} \emph {et~al.}}]{Aubert:2001nu}%
  \BibitemOpen
  \bibfield  {author} {\bibinfo {author} {\bibfnamefont {B.}~\bibnamefont {Aubert}} \emph {et~al.} (\bibinfo {collaboration} {BABAR Collaboration}),\ }\href {https://doi.org/10.1103/PhysRevLett.87.091801} {\bibfield  {journal} {\bibinfo  {journal} {Phys. Rev. Lett.}\ }\textbf {\bibinfo {volume} {87}},\ \bibinfo {pages} {091801} (\bibinfo {year} {2001})}\BibitemShut {NoStop}%
\bibitem [{\citenamefont {Abe}\ \emph {et~al.}(2001)\citenamefont {Abe} \emph {et~al.}}]{Abe:2001xe}%
  \BibitemOpen
  \bibfield  {author} {\bibinfo {author} {\bibfnamefont {K.}~\bibnamefont {Abe}} \emph {et~al.} (\bibinfo {collaboration} {Belle Collaboration}),\ }\href {https://doi.org/10.1103/PhysRevLett.87.091802} {\bibfield  {journal} {\bibinfo  {journal} {Phys. Rev. Lett.}\ }\textbf {\bibinfo {volume} {87}},\ \bibinfo {pages} {091802} (\bibinfo {year} {2001})}\BibitemShut {NoStop}%
\bibitem [{\citenamefont {Aaij}\ \emph {et~al.}(2019)\citenamefont {Aaij} \emph {et~al.}}]{Aaij:2019kcg}%
  \BibitemOpen
  \bibfield  {author} {\bibinfo {author} {\bibfnamefont {R.}~\bibnamefont {Aaij}} \emph {et~al.} (\bibinfo {collaboration} {LHCb Collaboration}),\ }\href {https://doi.org/10.1103/PhysRevLett.122.211803} {\bibfield  {journal} {\bibinfo  {journal} {Phys. Rev. Lett.}\ }\textbf {\bibinfo {volume} {122}},\ \bibinfo {pages} {211803} (\bibinfo {year} {2019})}\BibitemShut {NoStop}%
\bibitem [{\citenamefont {Aaij}\ \emph {et~al.}(2025)\citenamefont {Aaij} \emph {et~al.}}]{LHCb:2025ray}%
  \BibitemOpen
  \bibfield  {author} {\bibinfo {author} {\bibfnamefont {R.}~\bibnamefont {Aaij}} \emph {et~al.} (\bibinfo {collaboration} {LHCb Collaboration}),\ }\href {https://doi.org/10.1038/s41586-025-09119-3} {\bibfield  {journal} {\bibinfo  {journal} {Nature (London)}\ }\textbf {\bibinfo {volume} {643}},\ \bibinfo {pages} {1223} (\bibinfo {year} {2025})}\BibitemShut {NoStop}%
\bibitem [{\citenamefont {Tandean}(2004)}]{Tandean:2003fr}%
  \BibitemOpen
  \bibfield  {author} {\bibinfo {author} {\bibfnamefont {J.}~\bibnamefont {Tandean}},\ }\href {https://doi.org/10.1103/PhysRevD.69.076008} {\bibfield  {journal} {\bibinfo  {journal} {Phys. Rev. D}\ }\textbf {\bibinfo {volume} {69}},\ \bibinfo {pages} {076008} (\bibinfo {year} {2004})}\BibitemShut {NoStop}%
\bibitem [{\citenamefont {Salone}\ \emph {et~al.}(2022)\citenamefont {Salone}, \citenamefont {Adlarson}, \citenamefont {Batozskaya}, \citenamefont {Kupsc}, \citenamefont {Leupold},\ and\ \citenamefont {Tandean}}]{Salone:2022lpt}%
  \BibitemOpen
  \bibfield  {author} {\bibinfo {author} {\bibfnamefont {N.}~\bibnamefont {Salone}}, \bibinfo {author} {\bibfnamefont {P.}~\bibnamefont {Adlarson}}, \bibinfo {author} {\bibfnamefont {V.}~\bibnamefont {Batozskaya}}, \bibinfo {author} {\bibfnamefont {A.}~\bibnamefont {Kupsc}}, \bibinfo {author} {\bibfnamefont {S.}~\bibnamefont {Leupold}},\ and\ \bibinfo {author} {\bibfnamefont {J.}~\bibnamefont {Tandean}},\ }\href {https://doi.org/10.1103/PhysRevD.105.116022} {\bibfield  {journal} {\bibinfo  {journal} {Phys. Rev. D}\ }\textbf {\bibinfo {volume} {105}},\ \bibinfo {pages} {116022} (\bibinfo {year} {2022})}\BibitemShut {NoStop}%
\bibitem [{\citenamefont {Donoghue}\ \emph {et~al.}(1986)\citenamefont {Donoghue}, \citenamefont {He},\ and\ \citenamefont {Pakvasa}}]{Donoghue:1986hh}%
  \BibitemOpen
  \bibfield  {author} {\bibinfo {author} {\bibfnamefont {J.~F.}\ \bibnamefont {Donoghue}}, \bibinfo {author} {\bibfnamefont {X.-G.}\ \bibnamefont {He}},\ and\ \bibinfo {author} {\bibfnamefont {S.}~\bibnamefont {Pakvasa}},\ }\href {https://doi.org/10.1103/PhysRevD.34.833} {\bibfield  {journal} {\bibinfo  {journal} {Phys. Rev.}\ }\textbf {\bibinfo {volume} {D34}},\ \bibinfo {pages} {833} (\bibinfo {year} {1986})}\BibitemShut {NoStop}%
\bibitem [{\citenamefont {Lee}\ and\ \citenamefont {Yang}(1957)}]{Lee:1957qs}%
  \BibitemOpen
  \bibfield  {author} {\bibinfo {author} {\bibfnamefont {T.~D.}\ \bibnamefont {Lee}}\ and\ \bibinfo {author} {\bibfnamefont {C.-N.}\ \bibnamefont {Yang}},\ }\href {https://doi.org/10.1103/PhysRev.108.1645} {\bibfield  {journal} {\bibinfo  {journal} {Phys. Rev.}\ }\textbf {\bibinfo {volume} {108}},\ \bibinfo {pages} {1645} (\bibinfo {year} {1957})}\BibitemShut {NoStop}%
\bibitem [{\citenamefont {Ablikim}\ \emph {et~al.}(2024)\citenamefont {Ablikim} \emph {et~al.}}]{BESIII:2023jhj}%
  \BibitemOpen
  \bibfield  {author} {\bibinfo {author} {\bibfnamefont {M.}~\bibnamefont {Ablikim}} \emph {et~al.} (\bibinfo {collaboration} {BESIII Collaboration}),\ }\href {https://doi.org/10.1103/PhysRevLett.132.101801} {\bibfield  {journal} {\bibinfo  {journal} {Phys. Rev. Lett.}\ }\textbf {\bibinfo {volume} {132}},\ \bibinfo {pages} {101801} (\bibinfo {year} {2024})}\BibitemShut {NoStop}%
\bibitem [{\citenamefont {Ablikim}\ \emph {et~al.}(2022{\natexlab{a}})\citenamefont {Ablikim} \emph {et~al.}}]{BESIII:2021ypr}%
  \BibitemOpen
  \bibfield  {author} {\bibinfo {author} {\bibfnamefont {M.}~\bibnamefont {Ablikim}} \emph {et~al.} (\bibinfo {collaboration} {BESIII Collaboration}),\ }\href {https://doi.org/10.1038/s41586-022-04624-1} {\bibfield  {journal} {\bibinfo  {journal} {Nature (London)}\ }\textbf {\bibinfo {volume} {606}},\ \bibinfo {pages} {64} (\bibinfo {year} {2022}{\natexlab{a}})}\BibitemShut {NoStop}%
\bibitem [{\citenamefont {Adlarson}\ and\ \citenamefont {Kupsc}(2019)}]{Adlarson:2019jtw}%
  \BibitemOpen
  \bibfield  {author} {\bibinfo {author} {\bibfnamefont {P.}~\bibnamefont {Adlarson}}\ and\ \bibinfo {author} {\bibfnamefont {A.}~\bibnamefont {Kupsc}},\ }\href {https://doi.org/10.1103/PhysRevD.100.114005} {\bibfield  {journal} {\bibinfo  {journal} {Phys. Rev. D}\ }\textbf {\bibinfo {volume} {100}},\ \bibinfo {pages} {114005} (\bibinfo {year} {2019})}\BibitemShut {NoStop}%
\bibitem [{\citenamefont {Ablikim}\ \emph {et~al.}(2023)\citenamefont {Ablikim} \emph {et~al.}}]{Besiii:2023drj}%
  \BibitemOpen
  \bibfield  {author} {\bibinfo {author} {\bibfnamefont {M.}~\bibnamefont {Ablikim}} \emph {et~al.} (\bibinfo {collaboration} {BESIII Collaboration}),\ }\href {https://doi.org/10.1103/PhysRevD.108.L031106} {\bibfield  {journal} {\bibinfo  {journal} {Phys. Rev. D}\ }\textbf {\bibinfo {volume} {108}},\ \bibinfo {pages} {L031106} (\bibinfo {year} {2023})}\BibitemShut {NoStop}%
\bibitem [{\citenamefont {Huang}\ \emph {et~al.}(2017)\citenamefont {Huang}, \citenamefont {Zhang}, \citenamefont {Li},\ and\ \citenamefont {Kaiser}}]{Huang:2017bmx}%
  \BibitemOpen
  \bibfield  {author} {\bibinfo {author} {\bibfnamefont {B.-L.}\ \bibnamefont {Huang}}, \bibinfo {author} {\bibfnamefont {J.-S.}\ \bibnamefont {Zhang}}, \bibinfo {author} {\bibfnamefont {Y.-D.}\ \bibnamefont {Li}},\ and\ \bibinfo {author} {\bibfnamefont {N.}~\bibnamefont {Kaiser}},\ }\href {https://doi.org/10.1103/PhysRevD.96.016021} {\bibfield  {journal} {\bibinfo  {journal} {Phys. Rev. D}\ }\textbf {\bibinfo {volume} {96}},\ \bibinfo {pages} {016021} (\bibinfo {year} {2017})}\BibitemShut {NoStop}%
\bibitem [{\citenamefont {Holmstrom}\ \emph {et~al.}(2004)\citenamefont {Holmstrom} \emph {et~al.}}]{HyperCP:2004zvh}%
  \BibitemOpen
  \bibfield  {author} {\bibinfo {author} {\bibfnamefont {T.}~\bibnamefont {Holmstrom}} \emph {et~al.} (\bibinfo {collaboration} {HyperCP Collaboration}),\ }\href {https://doi.org/10.1103/PhysRevLett.93.262001} {\bibfield  {journal} {\bibinfo  {journal} {Phys. Rev. Lett.}\ }\textbf {\bibinfo {volume} {93}},\ \bibinfo {pages} {262001} (\bibinfo {year} {2004})}\BibitemShut {NoStop}%
\bibitem [{Note1()}]{Note1}%
  \BibitemOpen
  \bibinfo {note} {The HyperCP paper determined $\phi _{\Xi }$ and extracted the strong phase difference from $\alpha _{\Xi }\alpha _{\Lambda }$, relying on the assumed value $\alpha _{\Lambda }=0.642\pm 0.013$, which has since been revised by BESIII.}\BibitemShut {Stop}%
\bibitem [{\citenamefont {Nath}\ and\ \citenamefont {Kumar}(1965)}]{Nath:1965iud}%
  \BibitemOpen
  \bibfield  {author} {\bibinfo {author} {\bibfnamefont {R.}~\bibnamefont {Nath}}\ and\ \bibinfo {author} {\bibfnamefont {A.}~\bibnamefont {Kumar}},\ }\href {https://doi.org/10.1007/bf02751336} {\bibfield  {journal} {\bibinfo  {journal} {Nuovo Cimento}\ }\textbf {\bibinfo {volume} {36}},\ \bibinfo {pages} {669} (\bibinfo {year} {1965})}\BibitemShut {NoStop}%
\bibitem [{\citenamefont {Lu}\ \emph {et~al.}(1994)\citenamefont {Lu}, \citenamefont {Wise},\ and\ \citenamefont {Savage}}]{Lu:1994ex}%
  \BibitemOpen
  \bibfield  {author} {\bibinfo {author} {\bibfnamefont {M.}~\bibnamefont {Lu}}, \bibinfo {author} {\bibfnamefont {M.~B.}\ \bibnamefont {Wise}},\ and\ \bibinfo {author} {\bibfnamefont {M.~J.}\ \bibnamefont {Savage}},\ }\href {https://doi.org/10.1016/0370-2693(94)91456-7} {\bibfield  {journal} {\bibinfo  {journal} {Phys. Lett. B}\ }\textbf {\bibinfo {volume} {337}},\ \bibinfo {pages} {133} (\bibinfo {year} {1994})}\BibitemShut {NoStop}%
\bibitem [{\citenamefont {Kamal}(1998)}]{Kamal:1998se}%
  \BibitemOpen
  \bibfield  {author} {\bibinfo {author} {\bibfnamefont {A.~N.}\ \bibnamefont {Kamal}},\ }\href {https://doi.org/10.1103/PhysRevD.58.077501} {\bibfield  {journal} {\bibinfo  {journal} {Phys. Rev. D}\ }\textbf {\bibinfo {volume} {58}},\ \bibinfo {pages} {077501} (\bibinfo {year} {1998})}\BibitemShut {NoStop}%
\bibitem [{\citenamefont {Datta}\ \emph {et~al.}()\citenamefont {Datta}, \citenamefont {O'Donnell},\ and\ \citenamefont {Pakvasa}}]{Datta:1998pv}%
  \BibitemOpen
  \bibfield  {author} {\bibinfo {author} {\bibfnamefont {A.}~\bibnamefont {Datta}}, \bibinfo {author} {\bibfnamefont {P.}~\bibnamefont {O'Donnell}},\ and\ \bibinfo {author} {\bibfnamefont {S.}~\bibnamefont {Pakvasa}},\ }\href@noop {} {\ }\Eprint {https://arxiv.org/abs/hep-ph/9806374} {arXiv:hep-ph/9806374} \BibitemShut {NoStop}%
\bibitem [{\citenamefont {Datta}\ and\ \citenamefont {Pakvasa}(1995)}]{Datta_1995}%
  \BibitemOpen
  \bibfield  {author} {\bibinfo {author} {\bibfnamefont {A.}~\bibnamefont {Datta}}\ and\ \bibinfo {author} {\bibfnamefont {S.}~\bibnamefont {Pakvasa}},\ }\href {https://doi.org/10.1016/0370-2693(94)01577-y} {\bibfield  {journal} {\bibinfo  {journal} {Phys. Lett. B}\ }\textbf {\bibinfo {volume} {344}},\ \bibinfo {pages} {430} (\bibinfo {year} {1995})}\BibitemShut {NoStop}%
\bibitem [{\citenamefont {Tandean}\ \emph {et~al.}(2001)\citenamefont {Tandean}, \citenamefont {Thomas},\ and\ \citenamefont {Valencia}}]{Tandean:2000dx}%
  \BibitemOpen
  \bibfield  {author} {\bibinfo {author} {\bibfnamefont {J.}~\bibnamefont {Tandean}}, \bibinfo {author} {\bibfnamefont {A.~W.}\ \bibnamefont {Thomas}},\ and\ \bibinfo {author} {\bibfnamefont {G.~E.}\ \bibnamefont {Valencia}},\ }\href {https://doi.org/10.1103/PhysRevD.64.014005} {\bibfield  {journal} {\bibinfo  {journal} {Phys. Rev. D}\ }\textbf {\bibinfo {volume} {64}},\ \bibinfo {pages} {014005} (\bibinfo {year} {2001})}\BibitemShut {NoStop}%
\bibitem [{\citenamefont {Kaiser}(2001)}]{Kaiser:2001hr}%
  \BibitemOpen
  \bibfield  {author} {\bibinfo {author} {\bibfnamefont {N.}~\bibnamefont {Kaiser}},\ }\href {https://doi.org/10.1103/PhysRevC.73.069902} {\bibfield  {journal} {\bibinfo  {journal} {Phys. Rev. C}\ }\textbf {\bibinfo {volume} {64}},\ \bibinfo {pages} {045204} (\bibinfo {year} {2001})};\ \bibinfo {note} {\textbf{73}, 069902(E) (2006)}\BibitemShut {NoStop}%
\bibitem [{\citenamefont {Barros}()}]{Barros:2004pw}%
  \BibitemOpen
  \bibfield  {author} {\bibinfo {author} {\bibfnamefont {C.~C.}\ \bibnamefont {Barros}, \bibfnamefont {Jr.}},\ }\href@noop {} {\ }\Eprint {https://arxiv.org/abs/hep-ph/0402093} {arXiv:hep-ph/0402093} \BibitemShut {NoStop}%
\bibitem [{\citenamefont {Meissner}\ and\ \citenamefont {Oller}(2001)}]{Meissner:2000re}%
  \BibitemOpen
  \bibfield  {author} {\bibinfo {author} {\bibfnamefont {U.-G.}\ \bibnamefont {Meissner}}\ and\ \bibinfo {author} {\bibfnamefont {J.~A.}\ \bibnamefont {Oller}},\ }\href {https://doi.org/10.1103/PhysRevD.64.014006} {\bibfield  {journal} {\bibinfo  {journal} {Phys. Rev. D}\ }\textbf {\bibinfo {volume} {64}},\ \bibinfo {pages} {014006} (\bibinfo {year} {2001})}\BibitemShut {NoStop}%
\bibitem [{\citenamefont {Oller}\ \emph {et~al.}(2005)\citenamefont {Oller}, \citenamefont {Prades},\ and\ \citenamefont {Verbeni}}]{Oller:2005ig}%
  \BibitemOpen
  \bibfield  {author} {\bibinfo {author} {\bibfnamefont {J.~A.}\ \bibnamefont {Oller}}, \bibinfo {author} {\bibfnamefont {J.}~\bibnamefont {Prades}},\ and\ \bibinfo {author} {\bibfnamefont {M.}~\bibnamefont {Verbeni}},\ }\href {https://doi.org/10.1103/PhysRevLett.95.172502} {\bibfield  {journal} {\bibinfo  {journal} {Phys. Rev. Lett.}\ }\textbf {\bibinfo {volume} {95}},\ \bibinfo {pages} {172502} (\bibinfo {year} {2005})}\BibitemShut {NoStop}%
\bibitem [{\citenamefont {Oller}(2006)}]{Oller:2006jw}%
  \BibitemOpen
  \bibfield  {author} {\bibinfo {author} {\bibfnamefont {J.~A.}\ \bibnamefont {Oller}},\ }\href {https://doi.org/10.1140/epja/i2006-10011-3} {\bibfield  {journal} {\bibinfo  {journal} {Eur. Phys. J. A}\ }\textbf {\bibinfo {volume} {28}},\ \bibinfo {pages} {63} (\bibinfo {year} {2006})}\BibitemShut {NoStop}%
\bibitem [{\citenamefont {Guo}\ and\ \citenamefont {Oller}(2013)}]{Guo:2012vv}%
  \BibitemOpen
  \bibfield  {author} {\bibinfo {author} {\bibfnamefont {Z.-H.}\ \bibnamefont {Guo}}\ and\ \bibinfo {author} {\bibfnamefont {J.~A.}\ \bibnamefont {Oller}},\ }\href {https://doi.org/10.1103/PhysRevC.87.035202} {\bibfield  {journal} {\bibinfo  {journal} {Phys. Rev. C}\ }\textbf {\bibinfo {volume} {87}},\ \bibinfo {pages} {035202} (\bibinfo {year} {2013})}\BibitemShut {NoStop}%
\bibitem [{\citenamefont {Ablikim}\ \emph {et~al.}(2022{\natexlab{b}})\citenamefont {Ablikim} \emph {et~al.}}]{BESIII:2021cxx}%
  \BibitemOpen
  \bibfield  {author} {\bibinfo {author} {\bibfnamefont {M.}~\bibnamefont {Ablikim}} \emph {et~al.} (\bibinfo {collaboration} {BESIII Collaboration}),\ }\href {https://doi.org/10.1088/1674-1137/ac5c2e} {\bibfield  {journal} {\bibinfo  {journal} {Chin. Phys. C}\ }\textbf {\bibinfo {volume} {46}},\ \bibinfo {pages} {074001} (\bibinfo {year} {2022}{\natexlab{b}})}\BibitemShut {NoStop}%
\bibitem [{\citenamefont {Ablikim}\ \emph {et~al.}(2010)\citenamefont {Ablikim} \emph {et~al.}}]{BESIII:2009fln}%
  \BibitemOpen
  \bibfield  {author} {\bibinfo {author} {\bibfnamefont {M.}~\bibnamefont {Ablikim}} \emph {et~al.} (\bibinfo {collaboration} {BESIII Collaboration}),\ }\href {https://doi.org/10.1016/j.nima.2009.12.050} {\bibfield  {journal} {\bibinfo  {journal} {Nucl. Instrum. Methods Phys. Res., Sect. A}\ }\textbf {\bibinfo {volume} {614}},\ \bibinfo {pages} {345} (\bibinfo {year} {2010})}\BibitemShut {NoStop}%
\bibitem [{\citenamefont {Bai}\ \emph {et~al.}(2001)\citenamefont {Bai} \emph {et~al.}}]{BES:2001vqx}%
  \BibitemOpen
  \bibfield  {author} {\bibinfo {author} {\bibfnamefont {J.~Z.}\ \bibnamefont {Bai}} \emph {et~al.} (\bibinfo {collaboration} {BES Collaboration}),\ }\href {https://doi.org/10.1016/S0168-9002(00)00934-7} {\bibfield  {journal} {\bibinfo  {journal} {Nucl. Instrum. Methods Phys. Res., Sect. A}\ }\textbf {\bibinfo {volume} {458}},\ \bibinfo {pages} {627} (\bibinfo {year} {2001})}\BibitemShut {NoStop}%
\bibitem [{\citenamefont {Yu}\ \emph {et~al.}(2016)\citenamefont {Yu} \emph {et~al.}}]{Yu:2016cof}%
  \BibitemOpen
  \bibfield  {author} {\bibinfo {author} {\bibfnamefont {C.}~\bibnamefont {Yu}} \emph {et~al.},\ }in\ \href {https://doi.org/10.18429/JACoW-IPAC2016-TUYA01} {\emph {\bibinfo {booktitle} {{7th International Particle Accelerator Conference}}}}\ (\bibinfo {address} {Busan, Korea},\ \bibinfo {year} {2016}),\ p.\ \bibinfo {pages} {TUYA01}\BibitemShut {NoStop}%
\bibitem [{\citenamefont {Agostinelli}\ \emph {et~al.}(2003)\citenamefont {Agostinelli} \emph {et~al.}}]{GEANT4:2002zbu}%
  \BibitemOpen
  \bibfield  {author} {\bibinfo {author} {\bibfnamefont {S.}~\bibnamefont {Agostinelli}} \emph {et~al.} (\bibinfo {collaboration} {GEANT4 Collaboration}),\ }\href {https://doi.org/10.1016/S0168-9002(03)01368-8} {\bibfield  {journal} {\bibinfo  {journal} {Nucl. Instrum. Methods Phys. Res., Sect. A}\ }\textbf {\bibinfo {volume} {506}},\ \bibinfo {pages} {250} (\bibinfo {year} {2003})}\BibitemShut {NoStop}%
\bibitem [{\citenamefont {Jadach}\ \emph {et~al.}(2001)\citenamefont {Jadach}, \citenamefont {Ward},\ and\ \citenamefont {Was}}]{Jadach:2000ir}%
  \BibitemOpen
  \bibfield  {author} {\bibinfo {author} {\bibfnamefont {S.}~\bibnamefont {Jadach}}, \bibinfo {author} {\bibfnamefont {B.~F.~L.}\ \bibnamefont {Ward}},\ and\ \bibinfo {author} {\bibfnamefont {Z.}~\bibnamefont {Was}},\ }\href {https://doi.org/10.1103/PhysRevD.63.113009} {\bibfield  {journal} {\bibinfo  {journal} {Phys. Rev. D}\ }\textbf {\bibinfo {volume} {63}},\ \bibinfo {pages} {113009} (\bibinfo {year} {2001})}\BibitemShut {NoStop}%
\bibitem [{\citenamefont {Ping}(2008)}]{Ping:2008zz}%
  \BibitemOpen
  \bibfield  {author} {\bibinfo {author} {\bibfnamefont {R.-G.}\ \bibnamefont {Ping}},\ }\href {https://doi.org/10.1088/1674-1137/32/8/001} {\bibfield  {journal} {\bibinfo  {journal} {Chin. Phys. C}\ }\textbf {\bibinfo {volume} {32}},\ \bibinfo {pages} {599} (\bibinfo {year} {2008})}\BibitemShut {NoStop}%
\bibitem [{\citenamefont {Navas}\ \emph {et~al.}(2024)\citenamefont {Navas} \emph {et~al.}}]{ParticleDataGroup:2024cfk}%
  \BibitemOpen
  \bibfield  {author} {\bibinfo {author} {\bibfnamefont {S.}~\bibnamefont {Navas}} \emph {et~al.} (\bibinfo {collaboration} {Particle Data Group}),\ }\href {https://doi.org/10.1103/PhysRevD.110.030001} {\bibfield  {journal} {\bibinfo  {journal} {Phys. Rev. D}\ }\textbf {\bibinfo {volume} {110}},\ \bibinfo {pages} {030001} (\bibinfo {year} {2024})}\BibitemShut {NoStop}%
\bibitem [{\citenamefont {Chen}\ \emph {et~al.}(2000)\citenamefont {Chen}, \citenamefont {Huang}, \citenamefont {Qi}, \citenamefont {Zhang},\ and\ \citenamefont {Zhu}}]{Chen:2000tv}%
  \BibitemOpen
  \bibfield  {author} {\bibinfo {author} {\bibfnamefont {J.~C.}\ \bibnamefont {Chen}}, \bibinfo {author} {\bibfnamefont {G.~S.}\ \bibnamefont {Huang}}, \bibinfo {author} {\bibfnamefont {X.~R.}\ \bibnamefont {Qi}}, \bibinfo {author} {\bibfnamefont {D.~H.}\ \bibnamefont {Zhang}},\ and\ \bibinfo {author} {\bibfnamefont {Y.~S.}\ \bibnamefont {Zhu}},\ }\href {https://doi.org/10.1103/PhysRevD.62.034003} {\bibfield  {journal} {\bibinfo  {journal} {Phys. Rev. D}\ }\textbf {\bibinfo {volume} {62}},\ \bibinfo {pages} {034003} (\bibinfo {year} {2000})}\BibitemShut {NoStop}%
\bibitem [{\citenamefont {Perotti}\ \emph {et~al.}(2019)\citenamefont {Perotti}, \citenamefont {Fäldt}, \citenamefont {Kupsc}, \citenamefont {Leupold},\ and\ \citenamefont {Song}}]{Perotti:2018wxm}%
  \BibitemOpen
  \bibfield  {author} {\bibinfo {author} {\bibfnamefont {E.}~\bibnamefont {Perotti}}, \bibinfo {author} {\bibfnamefont {G.}~\bibnamefont {Fäldt}}, \bibinfo {author} {\bibfnamefont {A.}~\bibnamefont {Kupsc}}, \bibinfo {author} {\bibfnamefont {S.}~\bibnamefont {Leupold}},\ and\ \bibinfo {author} {\bibfnamefont {J.~J.}\ \bibnamefont {Song}},\ }\href {https://doi.org/10.1103/PhysRevD.99.056008} {\bibfield  {journal} {\bibinfo  {journal} {Phys. Rev.}\ }\textbf {\bibinfo {volume} {D99}},\ \bibinfo {pages} {056008} (\bibinfo {year} {2019})}\BibitemShut {NoStop}%
\bibitem [{\citenamefont {Ablikim}\ \emph {et~al.}(2022{\natexlab{c}})\citenamefont {Ablikim} \emph {et~al.}}]{BESIII:2022lsz}%
  \BibitemOpen
  \bibfield  {author} {\bibinfo {author} {\bibfnamefont {M.}~\bibnamefont {Ablikim}} \emph {et~al.} (\bibinfo {collaboration} {BESIII Collaboration}),\ }\href {https://doi.org/10.1103/PhysRevD.106.L091101} {\bibfield  {journal} {\bibinfo  {journal} {Phys. Rev. D}\ }\textbf {\bibinfo {volume} {106}},\ \bibinfo {pages} {L091101} (\bibinfo {year} {2022}{\natexlab{c}})}\BibitemShut {NoStop}%
\bibitem [{\citenamefont {Ablikim}\ \emph {et~al.}(2022{\natexlab{d}})\citenamefont {Ablikim} \emph {et~al.}}]{BESIII:2022qax}%
  \BibitemOpen
  \bibfield  {author} {\bibinfo {author} {\bibfnamefont {M.}~\bibnamefont {Ablikim}} \emph {et~al.} (\bibinfo {collaboration} {BESIII Collaboration}),\ }\href {https://doi.org/10.1103/PhysRevLett.129.131801} {\bibfield  {journal} {\bibinfo  {journal} {Phys. Rev. Lett.}\ }\textbf {\bibinfo {volume} {129}},\ \bibinfo {pages} {131801} (\bibinfo {year} {2022}{\natexlab{d}})}\BibitemShut {NoStop}%
\bibitem [{\citenamefont {Bondar}\ \emph {et~al.}(2013)\citenamefont {Bondar} \emph {et~al.}}]{Charm-TauFactory:2013cnj}%
  \BibitemOpen
  \bibfield  {author} {\bibinfo {author} {\bibfnamefont {A.~E.}\ \bibnamefont {Bondar}} \emph {et~al.} (\bibinfo {collaboration} {Charm-Tau Factory Collaboration}),\ }\href {https://doi.org/10.1134/S1063778813090032} {\bibfield  {journal} {\bibinfo  {journal} {Phys. At. Nucl.}\ }\textbf {\bibinfo {volume} {76}},\ \bibinfo {pages} {1072} (\bibinfo {year} {2013})}\BibitemShut {NoStop}%
\bibitem [{\citenamefont {Achasov}\ \emph {et~al.}(2024)\citenamefont {Achasov} \emph {et~al.}}]{Achasov:2023gey}%
  \BibitemOpen
  \bibfield  {author} {\bibinfo {author} {\bibfnamefont {M.}~\bibnamefont {Achasov}} \emph {et~al.},\ }\href {https://doi.org/10.1007/s11467-023-1333-z} {\bibfield  {journal} {\bibinfo  {journal} {Front. Phys. (Beijing)}\ }\textbf {\bibinfo {volume} {19}},\ \bibinfo {pages} {14701} (\bibinfo {year} {2024})}\BibitemShut {NoStop}%
\bibitem [{\citenamefont {Barucca}\ \emph {et~al.}(2021)\citenamefont {Barucca} \emph {et~al.}}]{PANDA:2020zwv}%
  \BibitemOpen
  \bibfield  {author} {\bibinfo {author} {\bibfnamefont {G.}~\bibnamefont {Barucca}} \emph {et~al.} (\bibinfo {collaboration} {PANDA Collaboration}),\ }\href {https://doi.org/10.1140/epja/s10050-021-00386-y} {\bibfield  {journal} {\bibinfo  {journal} {Eur. Phys. J. A}\ }\textbf {\bibinfo {volume} {57}},\ \bibinfo {pages} {154} (\bibinfo {year} {2021})}\BibitemShut {NoStop}%
\bibitem [{BEP()}]{BEPC:2025xni}%
  \BibitemOpen
  \href@noop {} {}\bibinfo {note} {\href{https://cstr.cn/31109.02.BEPC}{https://cstr.cn/31109.02.BEPC}}\BibitemShut {NoStop}%
\end{thebibliography}
